\documentclass[a4paper,alpha-refs]{eSpectra}

\journal{aej}

\usepackage{graphicx}
\usepackage{siunitx}
\usepackage[spanish, english]{babel} 
\usepackage{amssymb}
\usepackage{comment}

\usepackage{academicons}
\usepackage{aas-macros}
\usepackage{doi}
\usepackage{fontawesome5}
\usepackage[left]{lineno}

\definecolor{orcidlogocol}{rgb}{0.65, 0.807, 0.223}
\newcommand{\orcid}[1]{$\,$\href{https://orcid.org/#1}{\textcolor{orcidlogocol}{\faOrcid}}}
 
\title{Deep Learning for Point Spread Function Modeling in Cosmology}

\author[1,2,\authfn{1},]{Dayana Andrea Henao Arbel\'aez}
\author[3,2\authfn{2}]{Pierre-Fran\c{c}ois~L\'eget}
\author[4,5,2\authfn{3}]{Andrés A. Plazas Malagón}

    \affil[1]{{Universidad de Antioquia, Medell\'in, Colombia}}
    \affil[2]{{RECA Internship Program}}
    \affil[3]{{Department of Astrophysical Sciences, Princeton University, Princeton, NJ 08544, USA}}
    \affil[4]{SLAC National Accelerator Laboratory, Menlo Park, CA, USA}
    \affil[5]{Kavli Institute for Particle Astrophysics and Cosmology, Stanford University, Stanford, CA, USA}

\authnote{\authfn{1} Undergraduate Student - Physics, dayana.henaoa@udea.edu.co}
\authnote{\authfn{2} Associate Research Scholar (leget@astro.princeton.edu)}
\authnote{\authfn{3} Rubin Operations Scientist (plazas@stanford.edu)}

\papercat{Scientific Article}

\runningauthor{Henao Arbel\'aez et al.}

\jvolume{4}
\jnumber{1}
\jyear{2026}
\begin{document}
\begin{frontmatter}
\maketitle

\selectlanguage{english}
\begin{abstract}
\justifying

We present the development of a data-driven, AI-based model of the Point Spread Function (PSF) that achieves higher accuracy than the current state-of-the-art approach, ``PSF in the Full Field-of-View" (PIFF). PIFF is widely used in leading weak-lensing surveys, including the Dark Energy Survey (DES), the Hyper Suprime-Cam (HSC) Survey, and the Vera C. Rubin Observatory Legacy Survey of Space and Time (LSST). The PSF characterizes how a point source, such as a star, is imaged after its light traverses the atmosphere and telescope optics, effectively representing the “blurred fingerprint” of the entire imaging system. Accurate PSF modeling is essential for weak gravitational lensing analyses, as biases in its estimation propagate directly into cosmic shear measurements—one of the primary cosmological probes of the expansion history of the Universe and the growth of large-scale structure for dark energy studies. To address the limitations of PIFF, which constructs PSF models independently for each CCD and therefore loses spatial coherence across the focal plane, we introduce a deep-learning–based framework for PSF reconstruction. In this approach, an autoencoder is trained on stellar images obtained with the Hyper Suprime-Cam (HSC) of the Subaru Telescope and combined with a Gaussian process to interpolate the PSF across the telescope’s full field of view. This hybrid model captures systematic variations across the focal plane and achieves a reconstruction error of $3.4 \times 10^{-6}$ compared to PIFF's $3.7 \times 10^{-6}$, laying the foundation for integration into the LSST Science Pipelines.
\end{abstract}

\qquad\quad\textbf{Keywords:} Point Spread Function, Autoencoder, Gaussian Process

\selectlanguage{spanish} 
\begin{abstract}
\justifying
Presentamos un modelo de la Función de Dispersión de Punto (PSF, por sus siglas en inglés) basado en inteligencia artificial y guiado por datos, que supera en precisión al método actual de referencia, \textit{PSF in the Full Field-of-View} (PIFF). PIFF se utiliza ampliamente en los principales estudios de lente gravitacional débil, como el \textit{Dark Energy Survey} (DES), el \textit{Hyper Suprime-Cam} (HSC) Survey y la \textit{Investigación del Espacio-Tiempo como Legado para la Posteridad} (LSST) del Observatorio Vera C. Rubin. 
La PSF describe cómo se observa una fuente puntual, como una estrella, tras atravesar la atmósfera y la óptica del telescopio, representando la ``huella borrosa'' del sistema de imagen. Su modelado preciso es esencial para los análisis de lente gravitacional débil, ya que los sesgos en su estimación afectan directamente las mediciones de cizalladura cósmica, un método clave para estudiar la expansión del Universo y el crecimiento de la estructura a gran escala. Para superar las limitaciones de PIFF, que modela la PSF de forma independiente en cada CCD y pierde coherencia espacial en el plano focal, proponemos un marco de reconstrucción basado en aprendizaje profundo. Un autoencoder entrenado con imágenes estelares del \textit{Hyper Suprime-Cam} (HSC) del Telescopio Subaru se combina con un proceso gaussiano para interpolar la PSF en todo el campo de visión. Este modelo híbrido captura variaciones sistemáticas en el plano focal y reduce el error de reconstrucción a $3.4 \times 10^{-6}$, frente al $3.7 \times 10^{-6}$ de PIFF, sentando las bases para su futura implementación en las \textit{LSST Science Pipelines} del LSST.
\end{abstract}

\begin{skeywords}
Función de Dispersión de Punto, Autoencoder, Proceso Gaussiano
\end{skeywords}
\end{frontmatter}

\selectlanguage{english}

        \section{Introduction}
        
        Cosmology is the scientific discipline that, through observations of light and other cosmic signals such as gravitational waves, cosmic rays, and the cosmic microwave background, investigates the origin, structure, evolution, and ultimate fate of the Universe \cite{liddle2003introduction, bridle2009handbook}. The development of cosmology has been marked by key debates and discoveries that gradually defined its scientific foundations. \\

        A good example is the so-called “Great Debate” of 1920, in which Harlow Shapley and Heber Curtis discussed the scale of the cosmos. While Shapley argued that the Milky Way encompassed the entire Universe, Curtis maintained that spiral nebulae such as Andromeda were in fact independent galaxies \cite{Trimble1995}. This controversy was settled only a few years later, when Edwin Hubble measured the distance to Andromeda and revealed that the Universe extends far beyond the Milky Way. He also provided the first observational evidence for the expansion of the Universe, consistent with Georges Lemaître’s theoretical predictions from general relativity \cite{BritannicaRedshift2025}.\\

        Later, Fritz Zwicky noted that the visible matter in galaxy clusters was insufficient to explain their dynamics, suggesting the existence of “dark matter,” a hypothesis that decades later received strong support from Vera Rubin’s study of galaxy rotation curves. At the end of the 20th century, Saul Perlmutter, Adam Riess, and Brian Schmidt discovered through observations of Type Ia supernovae that cosmic expansion is accelerating, introducing the concept of “dark energy.” Despite decades of research, the fundamental nature of both dark matter and dark energy remains one of the most profound open questions in cosmology.\\
    
        As a result, modern cosmology recognizes that our Universe is dominated by dark energy, an unknown component driving the accelerated expansion of the cosmos, and dark matter, an invisible form of matter that interacts only through gravity. These components are not fully understood by modern science. In the case of dark matter, there is much evidence for its existence, suggesting that it may be a particle interacting gravitationally and, at most, weakly with other forces. However, it remains unknown which particle beyond the Standard Model could account for it. With respect to dark energy, its nature is entirely mysterious, and one of the most powerful observational probes to constrain it is weak gravitational lensing \cite{liddle2003introduction, bridle2009handbook, hobson2006general}. \\
           
        Gravitational lensing is the deflection of light rays from distant celestial objects caused by the curvature of spacetime induced by the presence of mass–energy distributions, analogous to the way light bends when it passes through different media. A gravitational lens is a localized concentration of mass and energy, such as a galaxy or a cluster of galaxies, that distorts and focuses the trajectory of light, similar to the action of an optical lens, producing observable effects such as changes in the apparent position, shape, or brightness of the background source  \cite{plazas2014lentes, liddle2003introduction, kilbinger2015cosmology, mandelbaum2018weak}. Because the lensing effect depends on the mass distribution of the deflector (the lens), a detailed study of gravitational lenses allows researchers to infer valuable information about the total mass of the lensing object, including both its visible and dark matter components \cite{mandelbaum2018weak}.\\

        \subsection{Weak Gravitational Lensing}
        
        Depending on the level of distortion produced in the observed image, gravitational lensing can be classified into different regimes. In extreme cases, known as strong gravitational lensing, multiple images of the same source or even complete Einstein rings can be observed. Conversely, when distortions are subtle and not directly visible in individual images, it is necessary to average over many background galaxies and apply statistical methods to detect the effect, a phenomenon referred to as weak gravitational lensing \cite{plazas2014lentes, mandelbaum2018weak}. \\

        Weak lensing is used to measure cluster masses and perform cosmological studies. An even subtler effect is cosmic shear, which provides key insights into the large-scale distribution of matter in the Universe \cite{kilbinger2015cosmology, amon2021desy3cosmicshear}.\\
        
        Cosmic shear is the subtle distortion in the shapes of galaxies. For most galaxies, this effect can be described as a linear transformation between unlensed $(x_u, y_u)$ and lensed $(x_l, y_l)$ coordinates applied to the entire galaxy image

        \[
        \begin{pmatrix}
        x_u \\ y_u
        \end{pmatrix}
        =
        \begin{pmatrix}
        1 - g_1  & -g_2 \\
        -g_2 & 1 + g_1
        \end{pmatrix}
        \begin{pmatrix}
        x_l \\ y_l
        \end{pmatrix}
        \]\\

    Here, a positive “shear” $g_1$ stretches an image along the $x$ axis and compresses along the $y$ axis; a positive shear $g_2$ stretches an image along the diagonal $y = x$ and compresses along $y = -x$ \cite{bridle2009handbook}. \\ 
      
    These distortions tend to align galaxy shapes perpendicular to the center of the dark matter overdensity that bends the light. Since this effect is extremely small, it cannot be detected in an individual galaxy; therefore, a statistical analysis of millions of galaxy shapes is required to uncover the lensing signal \cite{plazas2014lentes, amon2021desy3cosmicshear}. To extract significant results for cosmology, it is necessary to measure the cosmic shear with extremely high accuracy for millions of galaxies, despite observational challenges such as blurring, pixelisation, noise, and theoretical uncertainties regarding the undistorted shapes of galaxies \cite{mandelbaum2018weak}.\\
     
    To accurately measure the coherent distortions produced by weak gravitational lensing, it is essential to correct for the effects of the Point Spread Function (PSF). The PSF describes how a point source—such as a star—appears in an image after its light passes through the Earth's atmosphere, the telescope’s optics, and CCD detectors, including effects of pixelation. In other words, it represents the blurred image of an ideal point source and acts as a characteristic “fingerprint” of the entire imaging system. If the PSF is not properly accounted for, two main issues arise. First, the estimated shear values will tend to be underestimated, particularly for small galaxies. Second, the PSF can imprint additional distortions onto galaxy shapes, further biasing the measurement \cite{kilbinger2015cosmology}.\\

    An accurate PSF correction is important for reliable weak lensing measurements. Currently, the PSF is modeled using the PSFs in the Full FOV (PIFF) software package \cite{Jarvis2021}, which represents the state of the art in ground-based experiments for wide-field surveys such as DESC, HSC, or the Rubin Observatory. Although the name suggests modeling over the full focal plane, in practice the PSF used for cosmic shear analysis is obtained from PIFF using a per-CCD model, which has consistently provided better performance. Modern wide-field telescopes are composed of multiple CCDs, and constructing the PSF from a single CCD often loses valuable information available across the full field of view (FoV). Therefore, even though the original goal of PIFF was to model the PSF across the entire FoV, the per-CCD approach remains the most effective solution for weak lensing studies.\\

    PIFF provides flexible basis sets for PSF modeling (including pixel-based, shapelets, and Gaussian mixtures), and can construct models in either chip or sky coordinates, properly accounting for the image WCS (World Coordinate System, i.e. the transformation from pixel positions on the detector to celestial coordinates on the sky). PIFF interpolates the PSF across the FoV using various methods such as polynomials or Gaussian processes, and can perform fitting in both real and Fourier space. Additional features include the incorporation of optical aberration information and automated outlier rejection of unsuitable stars, etc \cite{JarvisPiffOverview}.\\

    The software is configured through a YAML file that typically includes three or four sections: (i) input, which specifies the image, the star catalog, and the correspondence between catalog columns and positional information (x,y or RA,Dec), as well as optional parameters such as sky subtraction; (ii) select, which defines how candidate stars are chosen, either from stellar properties or via a size–magnitude diagram, with all unflagged sources used if this section is omitted; (iii) psf, which defines both the PSF profile at a given location and the interpolation scheme across the field of view, optionally including outlier rejection; and (iv) output, which controls the products written to disk, ranging from the PSF model itself to diagnostic plots and statistical catalogs \cite{JarvisPiffTutorial}.\\
    
    In this project, the objective is to develop a more accurate model than PIFF for the Point Spread Function using deep learning techniques, particularly autoencoders.

    \section{Dataset}

    The dataset used in this project was collected using the Subaru Telescope, operated by the National Astronomical Observatory of Japan (NAOJ) in collaboration with international partners. The observations were carried out with the Hyper Suprime-Cam (HSC), a wide-field optical imaging camera with a 1.8 deg² field of view (1.5 degrees in diameter). Its optical system includes an 820 mm diameter first lens and a 1.65-meter-long lens barrel. The HSC is equipped with 116 Hamamatsu deep-depletion CCDs (each 2k × 4k), totaling 870 megapixels \cite{hsc_ssp}.\\

    The data were processed using the HSC version of the LSST Science Pipelines \cite{PSTN-019}, specifically the one described by Bosch et al. (2018) \cite{2018PASJ...70S...5B}. The resulting dataset contains a total of 2,787,956 stars observed through 404 visits. For each star, the dataset provides 17 features, including the observed image, the PSF model generated by PIFF, with both normalized so that the sum over all pixels is equal to one, the standard deviation and ellipticity components (both observed and modeled), spatial positions on the CCD and within the telescope’s field of view (FoV), brightness, detector ID, visit number, and filter band. These features are particularly relevant for PSF modeling because they include its precise location in the telescope’s focal plane.\\
    
\begin{figure}[h!]
        \centering
        \includegraphics[scale=0.15]{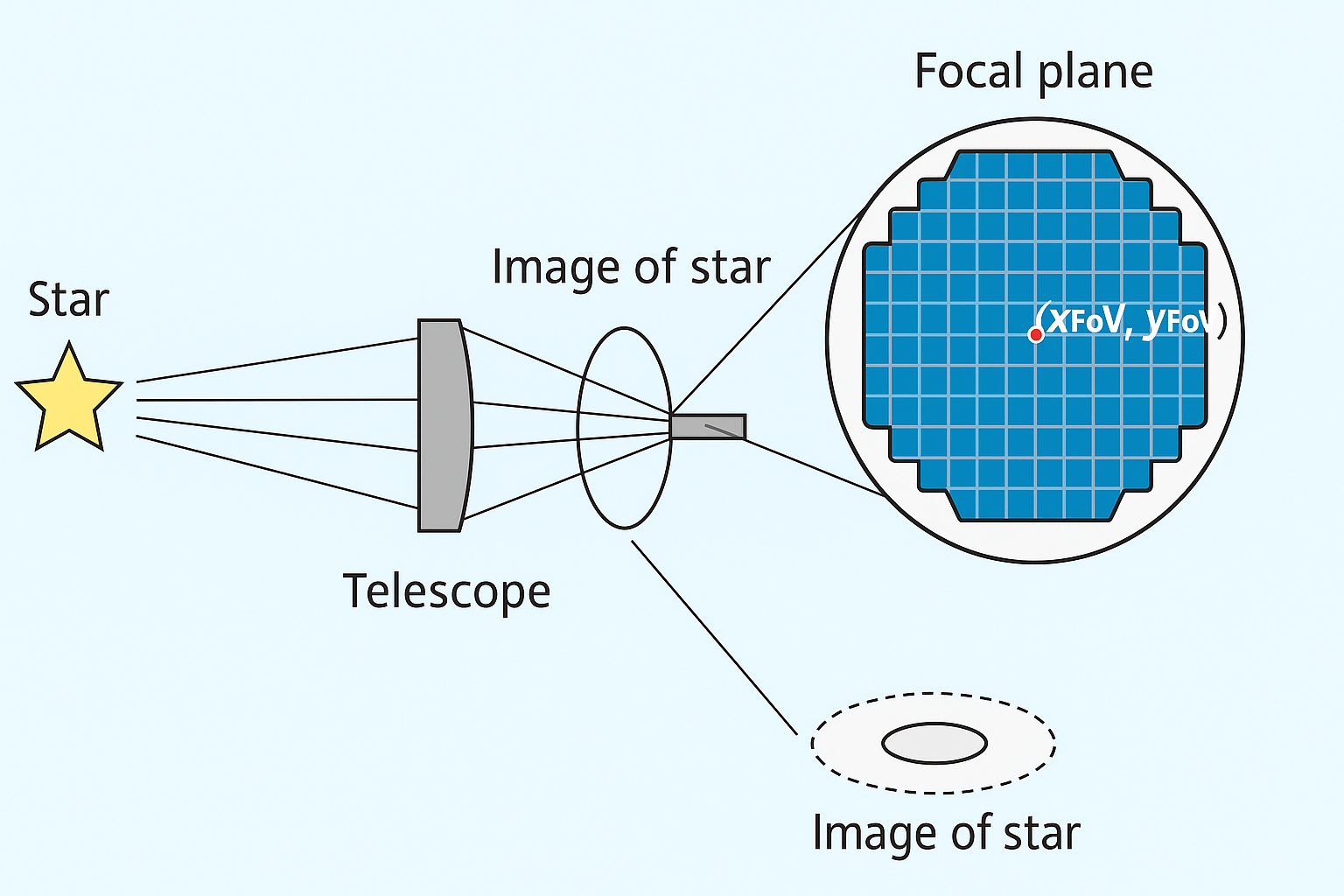}
        \caption{\label{fig:Telescopio}Description of the focal plane in the telescope.}
    \end{figure}

    As illustrated in Figure~\ref{fig:Telescopio}, the focal plane is where the starlight, after passing through the telescope’s optical system, converges to form an image. In large telescopes such as Subaru, the focal plane is a mosaic of many CCDs arranged together. Each star is recorded at a specific position, typically expressed in coordinates \texttt{xFoV} and \texttt{yFoV}. Since the PSF varies across the field of view, the basic idea of the PSF modeling process is to fit the PSF at the positions of stars and then use an interpolation scheme to propagate this model across the image. This allows one to estimate the PSF at the locations of galaxies, where weak lensing measurements are performed.\\
    
    A visit refers to a single exposure of the telescope, during which multiple CCDs capture different parts of the sky simultaneously. Traditionally, PSF models are constructed using data from a single CCD \cite{Jarvis2021}. However, for wide-field telescopes like Subaru, this approach can neglect valuable spatial information from the rest of the focal plane. Incorporating data from the full FoV allows for more accurate modeling, capturing systematic variations that occur across the instrument’s entire optical path. For this reason, an autoencoder was explored in this project for PSF modeling, and a Gaussian Process was employed to propagate the PSF across the Field of View.
    
    \section{Autoencoder and Gaussian Process}
    
    \subsection{Autoencoder}

    An autoencoder is a type of deep learning model whose objective is to compress high-dimensional data into a lower-dimensional representation (encoding) and then reconstruct the original input (decoding). This is implemented using artificial neural networks (ANNs), computational models inspired by the structure and function of the human brain, where information flows through interconnected units called neurons arranged in layers \cite{Nielsen2015, Goodfellow2016}.\\

    \begin{figure}[h!]
    \centering
    \includegraphics[scale=0.33]{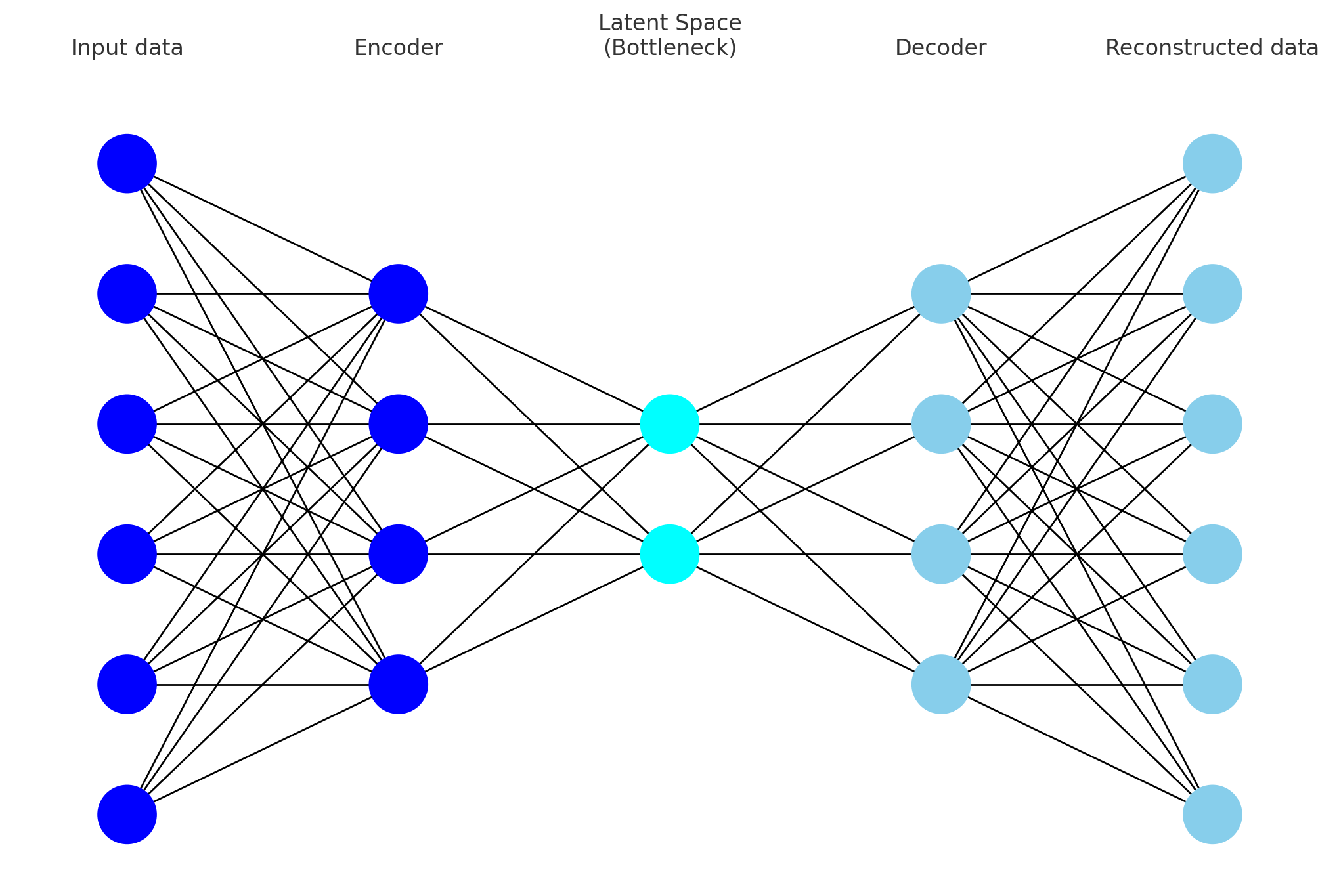}
    \caption{\label{fig:autoencoder} Architecture of the autoencoder.}
    \end{figure}

    A layer can be mathematically represented as a transformation of an input vector $\mathbf{x} \in \mathbb{R}^n$ into an output vector $\mathbf{y} \in \mathbb{R}^m$:
    \begin{equation}
        \mathbf{y} = f(\mathbf{W} \mathbf{x} + \mathbf{b}),
    \end{equation}
    where $\mathbf{W} \in \mathbb{R}^{m \times n}$ is the weight matrix, $\mathbf{b} \in \mathbb{R}^m$ is the bias vector, and $f(\cdot)$ is a non-linear activation function \cite{Nielsen2015, Goodfellow2016}.\\
    
    In each neuron, the weighted sum of the inputs is computed, a bias term is added, and the result is passed through the activation function. Non-linearity is crucial because it enables the network to capture complex patterns that cannot be represented by purely linear transformations. By stacking multiple layers, the model can learn progressively more abstract representations of the data, which in the case of an autoencoder correspond to the compressed encoding and its reconstruction. Without activation functions, this process becomes mathematically equivalent to a Principal Component Analysis (PCA), a statistical technique that reduces the dimensionality of data by finding new orthogonal axes, called principal components, that capture the maximum variance in the dataset. In this sense, an autoencoder can be interpreted as a non-linear generalization of PCA, where activation functions allow the model to capture more complex and non-linear relationships in the data \cite{Baldi1989, Bengio2013, Nielsen2015, Goodfellow2016}.\\

    The learning process consists of adjusting the weights and biases so that the predictions of the network approach the true values as closely as possible. This is achieved through a cycle of forward propagation, where data flows from the input to the output, loss computation, where the error between the prediction and the target is measured, and backpropagation, a method that updates the parameters by propagating the error backwards through the network. Optimization algorithms, such as stochastic gradient descent (SGD) or Adam, are typically used to perform these updates efficiently \cite{Nielsen2015, Goodfellow2016}.\\
    
    Several variables, known as hyperparameters, control the training process. Among them, the learning rate determines the size of the steps taken during parameter updates. The batch size specifies the number of training samples processed before the model’s parameters are updated \cite{Nielsen2015, Goodfellow2016}.\\
    
    The autoencoder consists of two main components: the encoder and the decoder, as illustrated in Fig.~\ref{fig:autoencoder}. The encoder maps the input data into a smaller latent representation, often referred to as the bottleneck, which captures the most relevant features of the data. The decoder then attempts to reconstruct the original input from this compressed representation \cite{kingma2013autoencoding, bergmann2023autoencoder}.
             
    The Autoencoder architecture implemented in this project consists of an encoder and a decoder composed of fully connected layers. The encoder first flattens the input images of size \(25 \times 25\) pixels (\(625\) elements) and progressively reduces their dimensionality through layers of \(312\), \(156\), and \(78\) neurons, each followed by a Rectified Linear Unit (ReLU) activation, defined as
    \begin{equation}
        g(z) = \max(0, z),
    \end{equation}
    which outputs the input value if it is positive and zero otherwise. This process continues until reaching the latent space of size \(16\).\\
    
    The decoder mirrors this structure, expanding the latent vector back to \(78\), \(156\), and \(312\) neurons, and finally to \(625\) elements with a Softmax activation to reconstruct the image. For a vector \(z \in \mathbb{R}^K\), the Softmax function is defined as
    \begin{equation}
        \text{softmax}(z)_i = \frac{e^{z_i}}{\sum_{j=1}^K e^{z_j}}, \quad i = 1, \dots, K,
    \end{equation}
    which converts a set of raw scores into a probability distribution over \(K\) possible classes \cite{Nielsen2015, Goodfellow2016}.  \\
    
    However, in this context, the Softmax activation ensures that the reconstructed images follow the same normalization as the input images, that is, that the sum of all pixel values in each image equals one. The reconstructed output is then reshaped to \(25 \times 25\) pixels. \\
    
    The loss function used in this work is the Mean Squared Error (MSE), which quantifies the difference between the predicted output $\hat{y}$ and the target output $y$. It is defined as
    \begin{equation}
        \text{MSE} = \frac{1}{n} \sum_{i=1}^{n} \left( \hat{y}_i - y_i \right)^2,
    \end{equation}
    where $n$ is the number of samples, $\hat{y}_i$ is the predicted value for the $i$-th sample, and $y_i$ is the corresponding target value \cite{Nielsen2015, Goodfellow2016}. \\
    
    The objective of an autoencoder is to minimize the reconstruction loss enabling the model to learn an efficient compression of the data while retaining as much relevant information as possible. In this case, the Adam optimizer was employed with a learning rate of \(1\times 10^{-3}\)  
    and a batch size of 128, which specifies the number of samples processed before each parameter update.\\
    
    Autoencoders are known to perform very well in minimizing mean squared error (MSE), making them effective for learning compact representations of the data. However, they do not learn a probabilistic distribution of the dataset; they only learn to reconstruct what they have already seen. This means that, while autoencoders achieve good reconstruction performance, they are not well-suited for generative tasks. If the decoder is fed with a random vector not coming from the encoder, it will not generate realistic samples but rather incoherent outputs or noise. This limitation motivated the introduction of Variational Autoencoders (VAEs), which incorporate a probabilistic framework that enables the generation of new, diverse samples. This topic will be discussed in more detail in the discussion section.

    \subsection{Gaussian Process Interpolation}

A Gaussian Process (GP) is a flexible and powerful method for interpolation, particularly well-suited for irregularly spaced datasets such as star observations across a telescope's focal plane. Mathematically, a GP models correlations between data points through a covariance function, commonly known as the kernel \cite{Leget2021}. This approach is also known as ordinary kriging in geostatistics and Wiener filtering in signal processing. While strictly speaking only the atmospheric contribution to the PSF can be considered a Gaussian random field, GPs are versatile enough to capture a wide range of features, including optical effects, which explains their widespread adoption in PSF modeling and related contexts.

\begin{figure*}
    \centering
    \includegraphics[width=\textwidth]{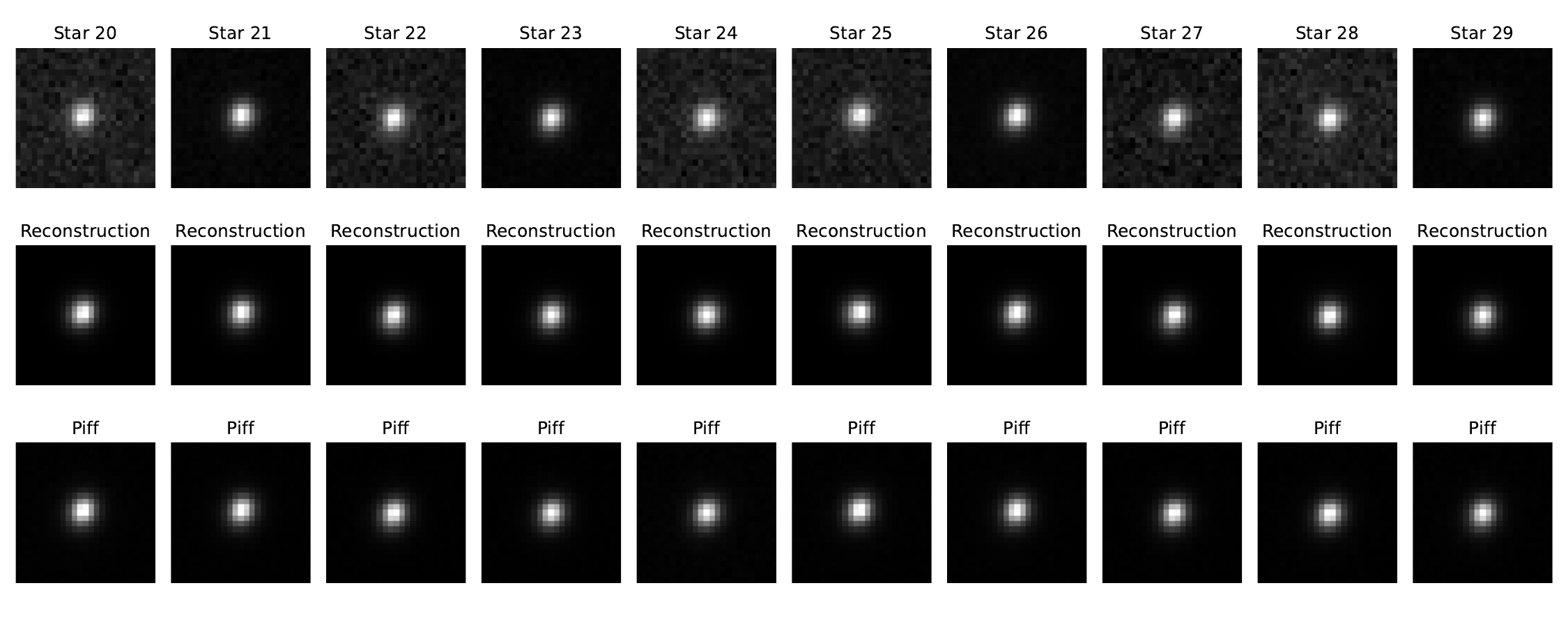}
    \captionof{figure}{Comparison between PSF reconstructions using the traditional PIFF model and the Autoencoder-based method (sample from the full dataset).}
    \label{fig:psf_comparison}
\end{figure*}
   
    Formally, a stationary Gaussian random field is fully characterized by its mean function and its covariance function, which depends only on the relative positions of data points. The kernel encodes these spatial correlations, and its analytical form must be chosen based on the expected properties of the data. The parameters of this kernel, often referred to as hyperparameters, describe the second-order statistics of the field, such as correlation length scales, amplitude, and anisotropy \cite{Rasmussen2006}.\\
    
    Given observations of the field at a set of positions, the GP interpolation provides the best linear unbiased prediction of the field at arbitrary locations, along with an estimate of the associated uncertainty. This capability is particularly valuable for modeling the spatially varying Point Spread Function across the focal plane, as it allows for continuous and smooth estimates based on discrete star measurements.\\
    
    In practice, the kernel parameters are estimated by maximizing the likelihood of the observed data under the GP model, which involves computations with the covariance matrix constructed from the kernel evaluated at observed positions. Efficient numerical techniques such as Cholesky decomposition and specialized packages like TreeCorr \cite{Leget2021} are employed to handle the large datasets typical in astronomical surveys.\\
    
    Overall, GP interpolation enables accurate reconstruction of PSF variations over the entire focal plane, capturing both isotropic and anisotropic spatial correlations without imposing restrictive assumptions on the functional form of the PSF variation.

    \section{Results}

Qualitatively, Figure~\ref{fig:psf_comparison} shows that the Autoencoder yields reconstructions that more closely resemble the original PSFs, particularly in terms of shape and symmetry. Quantitatively, the Mean Squared Error (MSE) computed between the reconstructed and original PSFs across the entire dataset confirms this result: the Autoencoder achieved an MSE of \(3.4\times 10^{-6}\), whereas PIFF obtained a slightly higher value of \(3.7\times 10^{-6}\). Thus, the Autoencoder not only produces visually more accurate reconstructions, but also achieves better numerical performance. \\

\begin{figure*}
    \centering
    \includegraphics[width=\textwidth]{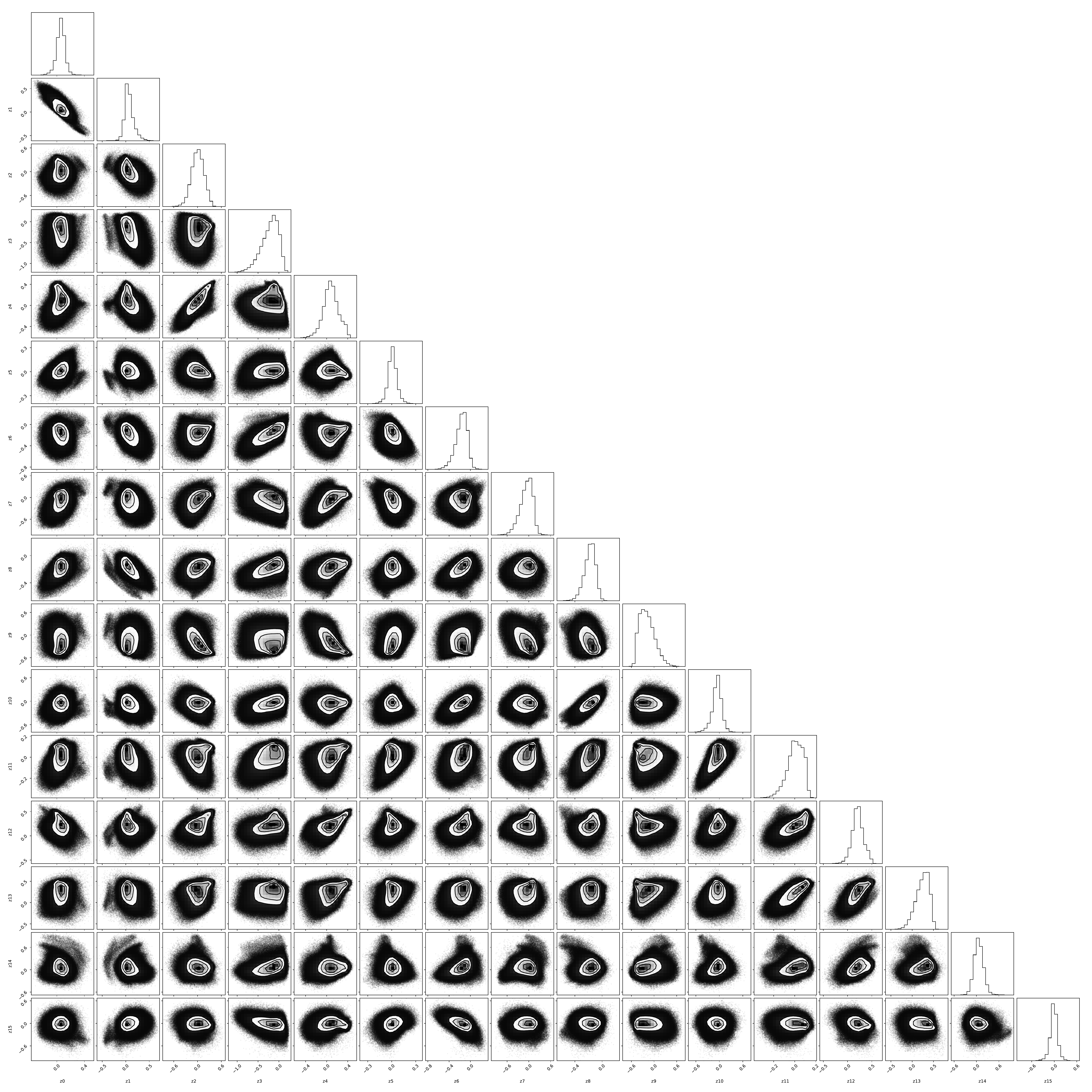}
    \captionof{figure}{Corner plot of the 16 latent dimensions of the autoencoder. The diagonal panels show the marginal distributions of each latent variable, while the off-diagonal panels display the pairwise correlations.}
    \label{fig:cornerplot}
\end{figure*}

Fig. \ref{fig:cornerplot} shows the correlations among the components of the latent space. Ideally, the latent variables should be independent, so that each dimension provides non-redundant information. However, the corner plot reveals correlations among several of them, indicating that the Autoencoder is not learning completely disentangled representations. This is not necessarily negative, since in this case the main objective is not to generate new images from the latent space, but rather to ensure precise reconstructions and spatial continuity in the latent representations across the FoV.\\

\begin{figure*}
    \centering
    \includegraphics[width=\textwidth]{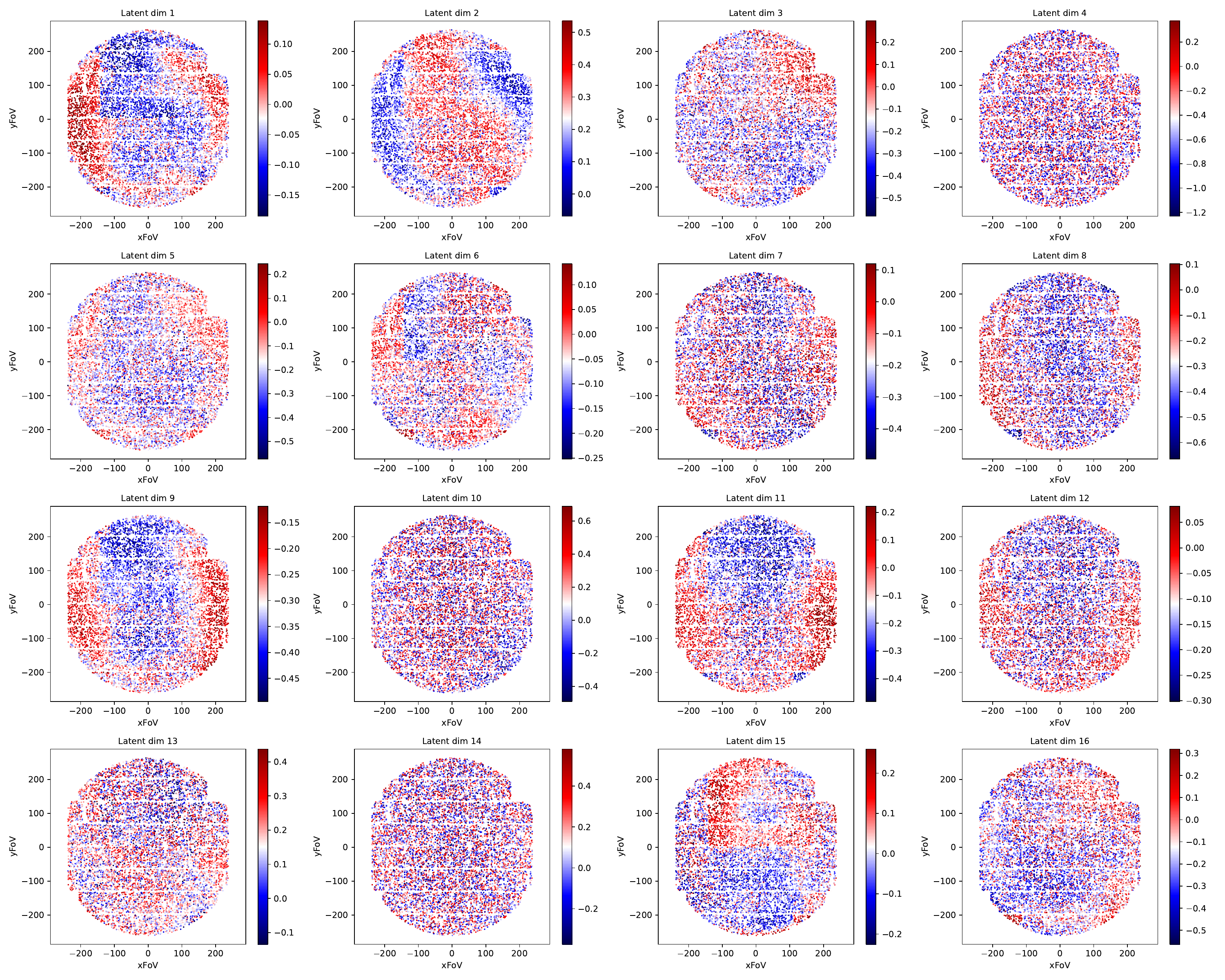}
    \captionof{figure}{Spatial distribution of the field of view (FoV) projected onto each of the latent dimensions of the autoencoder. Each panel shows how the corresponding latent variable captures variations of the PSF across the focal plane.}
    \label{fig:visita}
\end{figure*}

In Fig.~\ref{fig:visita}, the FoV corresponding to visit 38950 is shown, projected onto the 16 latent vectors generated by the Autoencoder. The latent components display smoothly varying structures across the focal plane, reflecting physical characteristics such as the telescope optics. Although Fig.~\ref{fig:cornerplot} does not show the latent correlations one might expect for generative AI applications, the absence of obvious discontinuities in the focal plane coordinates indicates that the observed patterns are consistent and physically meaningful.\\

\begin{figure*}
    \centering
    \includegraphics[width=\textwidth]{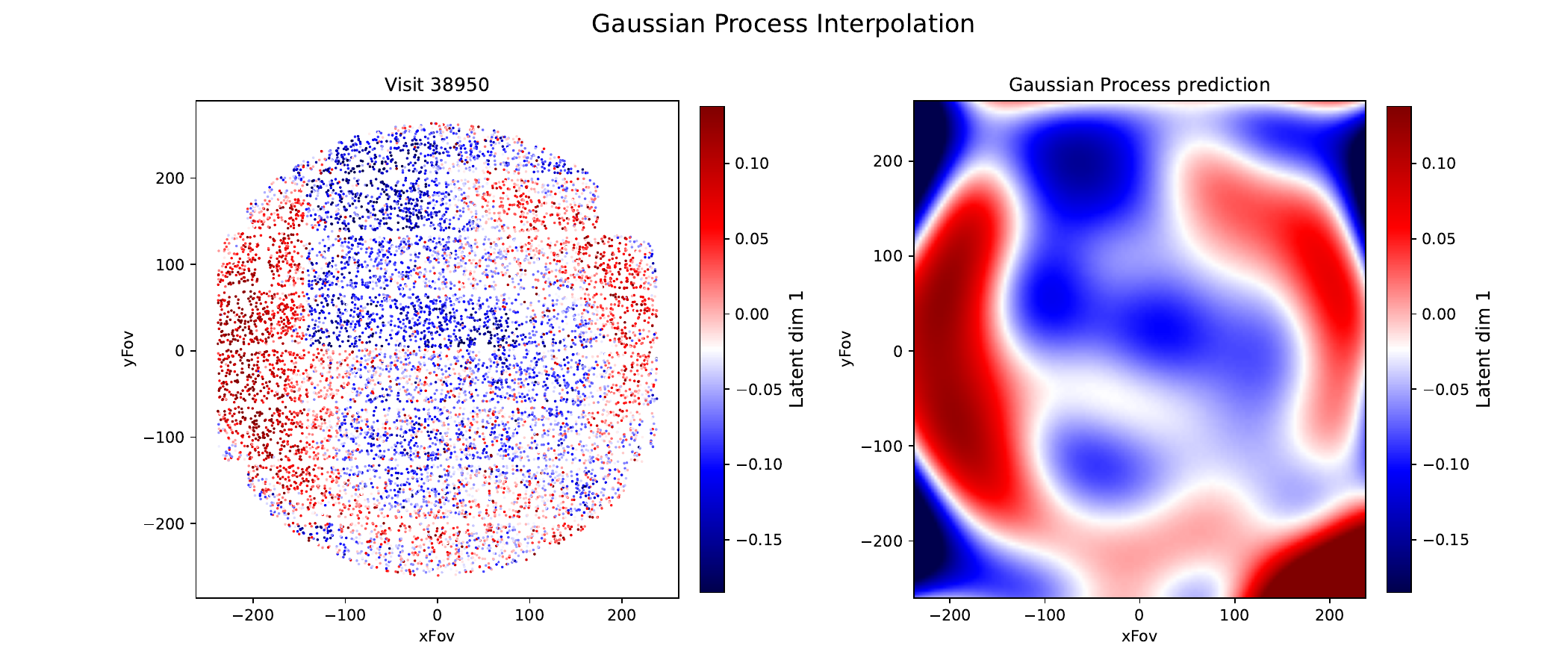}
    \captionof{figure}{Gaussian Process interpolation for the first latent dimension of the autoencoder. The left panel shows the observed values in the field of view, and the right panel displays the continuous prediction obtained from the Gaussian Process model.}
    \label{fig:GPI}
\end{figure*}

The interpolation performed using Gaussian Processes (Fig.~\ref{fig:GPI}) allowed for transforming the sparse distribution of the original data into a continuous map of the focal plane. The Gaussian Process was tested on a single visit and a single latent component, showing that it performs as expected and is able to capture the smooth variations across the focal plane.\\

\section{Conclusions and Discussions}

In this project, an autoencoder was trained on PSF stars from HSC data. The latent space representations learned by the model appear to outperform current state-of-the-art PSF modeling approaches. When projected onto the telescope focal plane, the latent variables capture physical variability across the field of view. By further applying Gaussian Processes to interpolate these latent features, it was shown that such variability can be consistently modeled across the focal plane.\\

This work effectively demonstrates a proof of concept: combining deep learning (via autoencoders) with Gaussian Processes can provide an alternative approach to PSF modeling. The next step toward full validation is to reconstruct PSFs by decoding the interpolated latent values and to test whether the performance still surpasses Piff on independent validation sets. Furthermore, this validation should be extended across multiple visits, rather than a single one, in order to assess robustness.\\

In general, Autoencoders present a trade-off between their reconstruction ability and their generative capacity. In this project, the priority was to optimize the reconstruction of stellar images, ensuring that the latent representations are consistent enough to capture PSF variations across the field of view and enable physically meaningful interpolations, which was successfully achieved. Nonetheless, future work will focus on implementing other types of autoencoders, such as Variational Autoencoders (VAEs) and Probabilistic Autoencoders (PAEs), which may be more suitable for data generation rather than reconstruction.\\

Building on these promising results, future work will include implementing this AI-based architecture within the LSST Science Pipelines as a dedicated measurement task. This integration will enable a direct comparison of both computational efficiency and modeling accuracy against the nominal PSF modeling task, which currently wraps the \texttt{Piff} framework. Such benchmarking within the Rubin Observatory data processing environment will be essential to assess the method’s scalability and readiness for production-level use. In parallel, future training will be extended to \textit{LSSTCam} data as the Rubin Observatory concludes its commissioning phase and the Legacy Survey of Space and Time (LSST) begins, allowing evaluation of the model’s performance under on-sky observing conditions and its potential for improving PSF modeling in early survey data releases.

\section{Acknowledgements}
The work of AAPM was supported by the U.S. Department of Energy under contract number DE-AC02-76SF00515.  AAPM  thanks the Department of Physics of Harvard University and the Laboratory of Particle Astrophysics and Cosmology, the Cosmology Group at Boston University, and the Department of Physics at Washington University in St. Louis for their hospitality during the preparation of this paper.
This research was conducted as part of the RECA (Network of Colombian Astronomy Students) Internship \citet{reca2023} Program 2025.




\bibliography{biblio}

\end{document}